\documentclass[aps,fleqn,preprint]{revtex4-1}

\usepackage{amsmath}
\usepackage{graphicx}
\usepackage{epsfig}
\usepackage{subfigure}
\usepackage{amsfonts}
\usepackage{amssymb}
\usepackage{amsthm}
\usepackage{newlfont}
\usepackage{epstopdf}

\usepackage{dcolumn}
\usepackage{bm}

\usepackage[section] {placeins}

\begin{document}


\title{A method for accurate electron-atom resonances:  
The complex-scaled multiconfigurational spin-tensor electron
propagator method for the $^2P\, \mbox{Be}^{-}$ shape resonance problem }

\author{Tsogbayar Tsednee, Liyuan Liang}
\author{Danny L.~Yeager}
\affiliation{Department of Chemistry, Texas A\&M University, College Station, Texas 77843, USA}

\begin{abstract}

We propose and develop the complex scaled multiconfigurational spin-tensor electron propagator (CMCSTEP) technique for theoretical determination 
of resonance parameters with electron-atom/molecule systems including open-shell and highly correlated atoms and molecules. The multiconfigurational spin-tensor electron propagator method (MCSTEP)
developed and implemented by Yeager his coworkers in real space gives very accurate and reliable ionization potentials and attachment energies.
The CMCSTEP method uses a complex scaled multiconfigurational self-consistent field (CMCSCF) state as an initial state along with a dilated Hamiltonian where all of the 
electronic coordinates are scaled by a complex factor. CMCSCF was developed and applied successfully to resonance problems earlier. We apply the CMCSTEP method to get $^2 P\,\mbox{Be}^{-}$ shape resonance parameters using $14s11p5d$, $14s14p2d$, and $14s14p5d$ basis sets with a $2s2p3d$\,CAS. The obtained value of the resonance parameters are compared to previous results. This is the first time CMCSTEP has been developed and used for a resonance problem. It will be among the most accurate and reliable techniques. 
Vertical ionization potentials and attachment energies in real space are typically within $\pm 0.2\,eV$ or better of excellent experiments and full configuration interaction calculations with a good basis set. We expect the same sort of agreement in complex space. 
\end{abstract}

\pacs{Valid PACS appear here}

\maketitle

\section{Introduction}

Resonances in electron-atom or -molecule scattering processes have attracted much attention. They play major roles in electron transport and energy exchange between electronic and nuclear motions, in vibrational excitation of molecules or molecular ions by electron impact, and dissociative attachments and recombination \cite{Brad, Mass}, and as a mechanism for DNA damage by low-energy electrons \cite{Barr,Sim}. 

In order to avoid direct calculation of an outgoing wave in resonance problems, we use a complex coordinate scaling (CS) technique, which was proposed and developed  by Aguilar, Balslev and Combes  \cite {Ag,Bal} and Simon \cite{Sim2} in the early 1970s. In this approach the electronic coordinates $(r)$ of the Hamiltonian are scaled (or 'dilated') by a complex parameter $\eta$ as $r\rightarrow \eta r$, where  $\eta = \alpha e^{i \theta}$ with $\alpha >0$ and $\theta \in (-\pi,\pi)$. Under this transformation, the bound states are real and are unchanged by complex scaling and the continua of the complex  scaled Hamiltonian $\bar{H}$ is rotated by an angle $2\theta$ at each threshold such that the continuum states appear as complex eigenvalues of the complex scaled Hamiltonian $\bar{H}$. The resonance parameters $E = E_{r} - i\frac{\Gamma_{r}}{2}$ hidden in the continua are exposed in complex space for some suitable $\eta$, where $E_{r}$ and $\Gamma_{r}$ are the resonance position and width of that resonance state, respectively. 

Other alternative methods have included the complex absorbing potential (CAP) \cite{Riss,Santra} instead of CS. CAP methods have not been shown conclusively to be superior to standard complex scaling. 

Previously, we developed the quadratically convergent complex scaled multiconfigurational self-consistent field \cite{Kous1,Kous2} (CMCSCF) method with step length control to obtain the resonance parameters. In real space, MCSCF with a small complete active space (CAS) has been proven to be a very effective method to describe nondynamical and some dynamical correlation correctly and is computationally cheaper than very large or full configuration interaction (CI) calculations \cite{Grah} while still incorporating the fundamental physics of what is going on. Based on the CMCSCF initial state, we also developed a new method termed as the $\mbox{M}_{1}$ method \cite{Kous2,Kous3}, in which the complex $\mbox{M}_{1}$ matrix is constructed from the first block of the $\mbox{M}$ matrix defined in MCSTEP \cite{Nicols1, Golab1, Yeager1,Yeagerdma,Yeagerdma3}. This block allows for only simple electron removal and addition to orbitals with no more complicated processes allowed to mix in.

MCSTEP, however, includes many additional operators which allow for more complicated electron ionization and attachment processes to be included. MCSTEP is designed to calculate reliably the ionization potentials (IPs) and attachment energies (AEs) for atoms and molecules which cannot generally be handled accurately by perturbation methods. In addition to simple electron addition operators to all orbitals as in the M1 method, MCSTEP includes operators the allow for electron removal and electron addition to all orbitals to excited states within the CAS \cite{Nicols1,Golab1,Yeager1,Yeagerdma,Yeagerdma3}.  In complex space, the $\mbox{M}_{1}$ and CMCSTEP methods use CMCSCF states as reference or initial state along with $\bar{H}$. Both the CMCSCF and $\mbox{M}_{1}$ methods have been previously efficiently used to study the $^2P\,\mbox{Be}^{-}$ shape resonance \cite{Kous1,Kous2,Kous3}.  

Moreover, we have developed and implemented the complex scaled multiconfigurational time-dependent Hartree-Fock method (CMCTDHF) [also called the complex scaled multiconfigurational linear response method (CMCLR)]. CMCTDHF uses CMCSCF state as the initial state. In real space multiconfigurational time-dependent Hartree-Fock (MCTDHF) has been successfully used to study electronic excitation energies and linear response properties \cite{Yeager2}. CMCTDHF has previously been implemented and successully employed to study Auger resonances  for $\mbox{Li}$ and $\mbox{Li}$-like cations \cite{Song1}, and $\mbox{Be}$ and $\mbox{Be}$-like cations \cite{Song2}  and  Feshbach resonances for both $\mbox{Be}^{+}(2p)$ \cite{Song3} and $\mbox{He}(2s^{2})$ \cite{Lly} systems, as well. 


In this work we implement the CMCSTEP method for the $^2P\,\mbox{Be}^{-}$ shape resonance problem using $14s11p5d$, $14s14p2d$, and $14s14p5d$ basis sets with a $2s2p3d$\,CAS and compare our results with previous results. The reasons why we implement this method for resonance problem are (i) MCSTEP in real space works exceptionally well and gives the most accurate and reliable values of vertical IPs and AEs for general atomic and molecular systems, which are well consistent with experimental measurements \cite{Grah2,YeagerN,Zak,Mah1,Mah2}, so that we expect that CMCSTEP is able to give the most reliable values of resonance parameters; (ii) even though this approach has been implemented for the first time  in complex space, this is a direct extension of the CMCSCF \cite{Kous1} and $\mbox{M}_{1}$ \cite{Kous2} methods which we previously developed and implemented, then we expect that the obtained results for this resonance problem will be different from those previously obtained and the most accurate \cite{Kous1,Kous2}.  


The paper is organized as follows. In section II we discuss the theoretical part of CMCSTEP method. In section III we present and discuss our results. Then conclusions follow.

\section{Theory}

The complex scaled electronic Hamiltonian is non-Hermitian. It is complex symmetric. This causes the wave function $|\psi_{m}\rangle$ to be complex conjugate biorthogonal (CCBON) where $\langle\psi^{\ast}_{i}|\psi_{j} \rangle=\delta_{ij}$ ($\ast$ means complex conjugate) \cite{Low}. It is shown that creation operators are introduced as $a^{T} = a^{\dagger} = (a^\ast)^{\dagger}$ rather than $a^{\dagger}$ with the usual anticommutation relations for creation and annihilation operators still hold by changing $"\dagger"$ into $"T"$ \cite{Yab,Yeager3}. 

Therefore, CMCSTEP may be formulated in the same way as MCSTEP via single particle Green's function or electron propagator method \cite{Nicols1, Golab1, Yeager1,Yeagerdma,Yeagerdma3} or super operator formalism \cite{Rowe} with the modified second quantization operators and $\bar{H}$.  We will not discuss MCSTEP in detail here, but they can be found in Refs.\cite{Nicols1, Golab1, Yeager1,Yeagerdma,Yeagerdma3}.  

CMCSTEP IPs and AEs are obtained from the following the complex generalized eigenvalue problem: 
\begin{equation}\label{cmcstepeq1}
 \mbox{{\bf M}} {\mathbf X}_{f} = \omega_{f} \mbox{{\bf N}} {\mathbf X}_{f}, 
\end{equation}
where
\begin{eqnarray}\label{cmcstepeq2}
 M_{rp} = \sum_{\Gamma} (-1)^{S_{0} -\Gamma - S_{f} - \gamma_{r}} W(\gamma_{r}\gamma_{p}S_{0}S_{0};\Gamma S_{f}) \nonumber\\
 \times (2\Gamma +1)^{1/2} \langle N S_{0}||\{ h^{\ast}_{r}(\bar{\gamma}_{r}),\bar{H},h_{p}(\gamma_{p})\}||NS_{0}\rangle,
\end{eqnarray}
and
\begin{eqnarray}\label{cmcstepeq3}
 N_{rp} = \sum_{\Gamma} (-1)^{S_{0} -\Gamma - S_{f} - \gamma_{r}} W(\gamma_{r}\gamma_{p}S_{0}S_{0};\Gamma S_{f}) \nonumber\\
\times (2\Gamma +1)^{1/2} \langle N S_{0}||\{ h^{\ast}_{r}(\bar{\gamma}_{r}),h_{p}(\gamma_{p})\}||NS_{0}\rangle,
\end{eqnarray}
$\omega_{f}$ is and IP or AE from the $N$-electron initial tensor state $|N S_{0}\rangle$ with spin $S_{0}$ to $N\pm 1$ electron final ion tensor state $|N\pm 1 S_{f}\rangle$ which has spin $S_{f}$. $W$ is the usual Racah coefficient, $h_{p}(\gamma_{p})$ and
$h^{\ast}_{p}(\bar{\gamma}_{r})$ are tensor operator versions of members of the operator manifold with ranks $\gamma_{p}$ and $\gamma_{r}$, respectively, $\{,\}$ is the anticommutator 
\begin{equation}\label{comm1}
 \{A,B\} = AB + BA,
\end{equation}
and $\{, ,\}$ is the symmetric double anticommutator 
\begin{equation}\label{comm2}
 \{A,B,C\} = \frac{1}{2}(\{A,[B,C]  + \{[A,B],C\}\} ).
\end{equation}
CMCSTEP uses a CMCSCF initial state with a fairly small CAS and couples tensor ionization and attachment operators to a tensor initial state to a final state that has the correct spin and spatial symmetry even if the initial state is open shell and/or highly correlated.

\section{Results and Discussion}

In this study, we investigate the low-lying $^2 P\,\mbox{Be}^{-}$ shape resonance problem using $\Delta$CMCSCF, $\mbox{M}_{1}$ and CMCSTEP methods. This resonance problem has been investigated theoretically in the past \cite{Ven,Donn,Kur1,Kur2,McCurdy,Resc,McN,Kryl1,Kryl2,Zhou,Langh}. 
Recently, we studied this resonance problem in terms of application of newly developed $\mbox{M}_{1}$ method \cite{Kous2}.  The $\mbox{Be}$ atom has a fairly  large amount of nondynamical correlation because the $1s^{2}2p^{2}$   
configuration has considerable mixing with the principle $1s^{2}2s^{2}$ configuration \cite{Golab1}, so that both configurations need to be included nonperturbatively for accurate IP and AE calculations. 

It is a common practice to report the resonance energy relative to the total energy of the scattering target. In this work, in $\Delta$CMCSCF calculations we report on the total energy of the continuum $\mbox{Be}^{-}$ species relative to that of $\mbox{Be}$ atom as 
\begin{equation}\label{cmcscfenfor}
  \epsilon_{{\tiny \Delta\mbox{CMCSCF}}} (\eta) = E^{N+1}_{c}  - E^{N}_{0},
\end{equation}
where, $E^{N+1}_{c}$ and $E^{N}_{0}$ are total energies of the $(N+1)$ electron $\mbox{Be}^{-}$ resonance state under investigation and the $N$ electron ground-state of the neutral $\mbox{Be}$ atom, respectively, and subscript $c$ and $0$ refer to continuum and bound states, respectively. 

In $\Delta$CMCSCF calculations we need to optimize each state separately, however, in $\mbox{M}_{1}$ and CMCSTEP calculations we can obtain energies of all states simultaneously.  In order to be consistent with the $\Delta$CMCSCF calculation,  we report on resonance parameter $ \epsilon_{{\tiny \mbox{CMCSTEP}}} $ obtained from CMCSTEP method: 
\begin{equation}\label{mcstepenfor}
  \epsilon_{{\tiny \mbox{CMCSTEP}}} (\eta) = \omega^{{\tiny \mbox{CMCSTEP}}} _{f} + E^{N}_{c}  - E^{N}_{0}, 
\end{equation}
where $\omega^{{\tiny \mbox{CMCSTEP}}} _{f} \equiv \omega_{f}$ is calculated from equation (\ref{cmcstepeq1}). In the case of $\mbox{M}_{1}$ calculations it is obtained from the $\mbox{M}_{1}$ complex eigenvalue problem \cite{Kous2} and we reported on results based on complex eigenvalues $\omega^{\tiny \mbox{M}_{1}}_{f}$ rather than $\omega^{{\tiny \mbox{CMCSTEP}}} _{f}$ in equation (\ref{mcstepenfor}). 

For this resonance problem, Venkatnathan et al \cite{Ven} found the $14s11p$ basis set to be the best one. We subsequently have shown that this $14s11p$ basis set is somewhat inadequate for resonances and that at least $14p$ functions are much more reliable. Hence, we chose this basis set initially and added $p$ and $d$ functions to it using a geometric progression with a view to account for the diffuse nature of the resonances.  Although for the very accurate IPs and excitation energies of $\mbox{Be}$, a larger $2s2p3s3p3d$\,CAS which enables more correlation is necessary \cite{Yeager1}, we employ a $2s2p3d$\,CAS with basis sets $14s11p5d$, $14s14p2d$ and $14s14p5d$ in this calculation, since we have previously found that a larger CAS is unnecessary for accurate shape resonance calculations \cite{Kous3}. However, most of the IP basis sets are designed for IPs where tighter functions are necessary rather than for resonance calculations where what is needed are basis functions to describe the near continuum. First, we performed $\Delta$CMCSCF calculations with all basis sets, and then followed it up with $\mbox{M}_{1}$ and CMCSTEP methods. The first two methods have already been implemented for the resonance problem with other basis sets \cite{Kous1,Kous2}; however, here the CMCSTEP method is applied for the first time for this resonance problem. Of these, CMCSTEP should be the most accurate, efficient and reliable method. So far, there are no experimental results for resonance parameters of $\mbox{Be}^{-}$.

In Table I we present IP for the $X\,^2S$ state of $\mbox{Be}$ atom and AE for $^2P\,\mbox{Be}^{-}$  resonance state obtained from $\Delta$CMCSCF, $\mbox{M}_{1}$ and MCSTEP ($\theta = 0\,rad$ and $\alpha=1$) calculations, in which $\mbox{Im(E) = 0}$. A comparison of values of IPs and AEs to previously obtained theoretical values and experimental measurements presented in this table show  that MCSTEP calculations are much better than MCSCF and $\mbox{M}_{1}$ approaches, and will give more reliable values for resonance problems. 

\begin{table}[h]
\caption{Ionization potential and attachment energies for $X ^2S$ of $\mbox{Be}$ atom and $^2P\,\mbox{Be}^{-}$ ion, respectively. }
\begin{center}
\begin{tabular}{c@{\hspace{2mm}}c@{\hspace{2mm}}c@{\hspace{2mm}}c@{\hspace{2mm}}c@{\hspace{2mm}}c@{\hspace{2mm}}c@{\hspace{2mm}}c@{\hspace{2mm}}c@{\hspace{2mm}}c@{\hspace{2mm}}c@{\hspace{2mm}}c@{\hspace{2mm}}c@{\hspace{2mm}}c@{\hspace{2mm}} }
\hline\hline
$\mbox{Method and basis set} $ &  $\mbox{IP}\,(eV)$ & $\mbox{AE}\,(eV)$ \\
\hline
$\Delta$MCSCF: $14s11p5d - 2s2p3d$\,CAS & 8.478  & 1.027 & \\
$\Delta$MCSCF: $14s14p2d - 2s2p3d$\,CAS & 8.042 & 1.124 & \\
$\Delta$MCSCF: $14s14p5d - 2s2p3d$\,CAS & 8.042 & 1.121& \\
\hline
$\mbox{M}_{1}$: $14s11p5d - 2s2p3d$\,CAS &  7.571 & 0.836 & \\
$\mbox{M}_{1}$: $14s14p2d - 2s2p3d$\,CAS &  7.576 & 0.872 & \\
$\mbox{M}_{1}$: $14s14p5d - 2s2p3d$\,CAS &  7.576 & 0.871 & \\
\hline
MCSTEP: $14s11p5d - 2s2p3d$\,CAS & 9.508 & 0.843 & \\ 
MCSTEP: $14s14p2d - 2s2p3d$\,CAS & 9.506 & 0.918 & \\ 
MCSTEP: $14s14p5d - 2s2p3d$\,CAS & 9.506 & 0.906 & \\
\hline
Ref.\cite{Golab1} $- 2s2p$\,CAS &  9.50 &  & \\ 
Ref.\cite{Golab1}$-2s2p3s3p3d$\,CAS &  9.31 &  & \\ 
Expt.\cite{Bash} &  9.32&  & \\ 
\hline\hline
\end{tabular} 
\end{center}
\end{table}

In Table II we show a summary of the obtained values of $^2P\,\mbox{Be}^{-}$ shape resonance for three different basis sets. In rows 2-4 of Table II we show results form $\Delta$CMCSCF calculations. These give larger widths than $\mbox{M}_{1}$ or CMCSTEP. Resonance parameters obtained from $\mbox{M}_{1}$ and CMCSTEP methods shown in rows 5-10 are fairly consistent with each other, although CMCSTEP will be more accurate. The optimal values of $\alpha$ and $\theta$ enables one to estimate the resonance parameters, and can be found by the system of equations  below:
 \begin{equation}\label{alftheta1}
  \frac{\partial E}{\partial \alpha} = \frac{\eta}{\alpha}\frac{\partial E}{\partial \eta} = 0 , 
\end{equation}
 \begin{equation}\label{alftheta2}
  \frac{\partial E}{\partial \theta} = -i\eta\frac{\partial E}{\partial \eta} = 0,
\end{equation}
which form the trajectory method by determining $E(\alpha_{opt},\theta_{opt})$ corresponding to the stability (loops, kinks, inflections, or any kind of "slow down") in the plots of $\mbox{Im}(E)$ as a function of $\mbox{Re}(E)$ evaluated as a series of $\alpha$ ($\alpha$ trajectory) and a series of $\theta$ values ($\theta$ trajectory) \cite{Donn}. 

Vertical IPs and AEs in real space are typically within $\pm 0.2\,eV$ of excellent experiments and full configuration interaction calculations with a good basis set \cite{Golab1}. We expect the same sort of agreement in complex space. Indeed this can be seen by looking at Table II  where our results are in general agreement with considerably less accurate methods.

\begin{table*}[h]
\caption{ Summary of theoretical calculations for $^2P\,\mbox{Be}^{-}$ shape resonance relative to $1s^22s^2$ ground state.  }
\begin{center}
{\scriptsize
\begin{tabular}{c@{\hspace{2mm}}c@{\hspace{2mm}}c@{\hspace{2mm}}c@{\hspace{2mm}}c@{\hspace{2mm}}c@{\hspace{2mm}}c@{\hspace{2mm}}c@{\hspace{2mm}}c@{\hspace{2mm}}c@{\hspace{2mm}}c@{\hspace{2mm}}c@{\hspace{2mm}}c@{\hspace{2mm}}c@{\hspace{2mm}} }
\hline\hline
$\mbox{Method and basis set} $ & $\alpha_{opt}$ & $\theta_{opt} (rad)$  &  $E_{r}\,(eV)$ & $\Gamma_{r}\,(eV)$ \\
\hline
$\Delta$CMCSCF: $14s11p5d - 2s2p3d$\,CAS & 1 & 0.49 & 0.714 & 1.541 & \\
$\Delta$CMCSCF: $14s14p2d - 2s2p3d$\,CAS & 1 & 0.55 & 0.816 & 1.731 & \\
$\Delta$CMCSCF: $14s14p5d - 2s2p3d$\,CAS & 1 & 0.55 & 0.819 & 1.736 & \\
\hline
$\mbox{M}_{1}$: $14s11p5d - 2s2p3d$\,CAS & 1 & 0.36 & 0.764 & 0.796 & \\
$\mbox{M}_{1}$: $14s14p2d - 2s2p3d$\,CAS & 1.03 & 0.36 & 0.790 & 0.856 & \\
$\mbox{M}_{1}$: $14s14p5d - 2s2p3d$\,CAS & 1.03 & 0.37 & 0.789 & 0.874 & \\
\hline
CMCSTEP: $14s11p5d - 2s2p3d$\,CAS &1 & 0.30 & 0.768 & 0.740 & \\
CMCSTEP: $14s14p2d - 2s2p3d$\,CAS &1.03 & 0.30 & 0.795 & 0.681 & \\
CMCSTEP: $14s14p5d - 2s2p3d$\,CAS &1.03 & 0.36 & 0.756 & 0.862 & \\
\hline\hline
\end{tabular} }
\end{center}
\end{table*}

In Table III we have listed theoretical results obtained by other workers. Our current results with our best basis set ($14s14p5d$) are quite far from complex CI \cite{McN} and density functional theory (DFT) combined with a CAP \cite{Zhou} calculations. We note that the CI calculations did not include any effect of quadruple excitations. It is well known that these need to be included for accurate CI energies and properties \cite{Langh}. The complex DFT calculation contains parameters that are experimentally determined and also the basis set used is small (contracted Gaussian $5s4p1d$ functions) and not adequate for resonance calculations. However, the CMCSTEP obtained values in this work are fairly comparable with those obtained by $\Delta$SCF \cite{McCurdy,Resc} and electron propagator methods \cite{McCurdy,Resc,Langh,Ven}, although these other methods will not be as accurate, since they are based on a single configuration and $\mbox{Be}$ atom is inherently multiconfigurational with the $1s^{2}2p^{2}$ configuration mixing in strongly (i.e.\,$10\%$) with the $1s^{2}2s^{2}$ configuration for the initial state. From our previous experience with calculations for IPs and AEs for atomic systems \cite{Golab1,Grah2,YeagerN}, we know that MCSTEP works very well, therefore we can say that value of resonance parameters obtained from CMCSTEP in this work are reliable.  

\begin{table*}[h]
\caption{Theoretical calculations for $^2P\,\mbox{Be}^{-}$ shape resonance.  }
\begin{center}
{\scriptsize
\begin{tabular}{c@{\hspace{2mm}}c@{\hspace{2mm}}c@{\hspace{2mm}}c@{\hspace{2mm}}c@{\hspace{2mm}}c@{\hspace{2mm}}c@{\hspace{2mm}}c@{\hspace{2mm}}c@{\hspace{2mm}}c@{\hspace{2mm}}c@{\hspace{2mm}}c@{\hspace{2mm}}c@{\hspace{2mm}}c@{\hspace{2mm}} }
\hline\hline
$\mbox{Method} $ &  &  $E_{r}\,(eV)$ & $\Gamma_{r}\,(eV)$ \\
\hline
Static exchange phase shift \cite{Kur1} & & 0.77 & 1.61 & \\
Static exchange phase shift plus polarizability phase shift \cite{Kur1} & & 0.20 & 0.28 & \\
Static exchange cross-section \cite{Kur2} & & 1.20 & 2.60 & \\
Static exchange plus polarizability cross-section \cite{Kur2} & & 0.16 & 0.14 & \\
$\Delta \mbox{SCF}$ with complex $14s16p$ Gaussian basis set \cite{McCurdy} & & 0.70 & 0.51 & \\
$\Delta \mbox{SCF}$ with complex $5s11p$ (Slater-type) basis set \cite{Resc} & & 0.76 & 1.11 & \\
Single, doubles, and triples complex CI \cite{McN} & & 0.32 & 0.30 & \\
$S$ matrix pole $(X_{\alpha})$ \cite{Kryl1,Kryl2} & & 0.10 & 0.15 & \\
Complex density functional theory \cite{Zhou} & & 0.580 & 0.223 & \\
Second-order dilated electron propagator based on real SCF \cite{Langh} & & 0.57 & 0.99 & \\
Bi-orthogonal dilated electron propagator (bases set $14s11p$ ) \cite{Ven}: & &  &  & \\
Zeroth order  & & 0.62 & 1.00 & \\
Quasiparticle second order  & & 0.61 & 1.00 & \\
Second order  & & 0.48 & 0.82 & \\
Quasiparticle third order  & & 0.54 & 0.82 & \\
OVGF third order  & & 0.54 & 0.78 & \\
Third order  & & 0.53 & 0.85 & \\
$\Delta$CMCSCF ($14s11p-2s2p3s3p$\,CAS) \cite{Kous1}  & & 0.73 & 1.58 & \\
$\mbox{M}_{1}$ ($14s11p3d-2s2p3s3p3d$\,CAS) \cite{Kous2}  & & 0.72 & 1.12 & \\
This work:  & &  &  & \\
$\Delta$CMCSCF ($14s14p5d-2s2p3d$\,CAS)  & & 0.819 & 1.736 & \\
$\mbox{M}_{1}$  ($14s14p5d-2s2p3d$\,CAS)    & & 0.789 & 0.874 & \\
CMCSTEP ($14s14p5d-2s2p3d$\,CAS)    & & 0.756 & 0.862 & \\
\hline\hline
\end{tabular} }
\end{center}
\end{table*}

In Figure 1 we show the $\theta$-trajectories for $^2P\,\mbox{Be}^{-}$ shape resonance obtained from the CMCSTEP method. Curves shown in panels (a), (b) and (c) are corresponded to calculations with basis sets $14s11p5d$, $14s14p2d$ and $14s14p5d$ with $2s2p3d$\,CAS, respectively. Crosses on each trajectory show a stabilized point. All trajectories show resonance points clearly along with an increased density of points.
In all trajectories $\theta$ starts at $\theta = 0.01$\,rad at the top and increases to down with a step of 0.01\,rad. 

\begin{figure*}[h]
\centering
    \mbox{\subfigure[]{\includegraphics[width=0.320\textwidth]{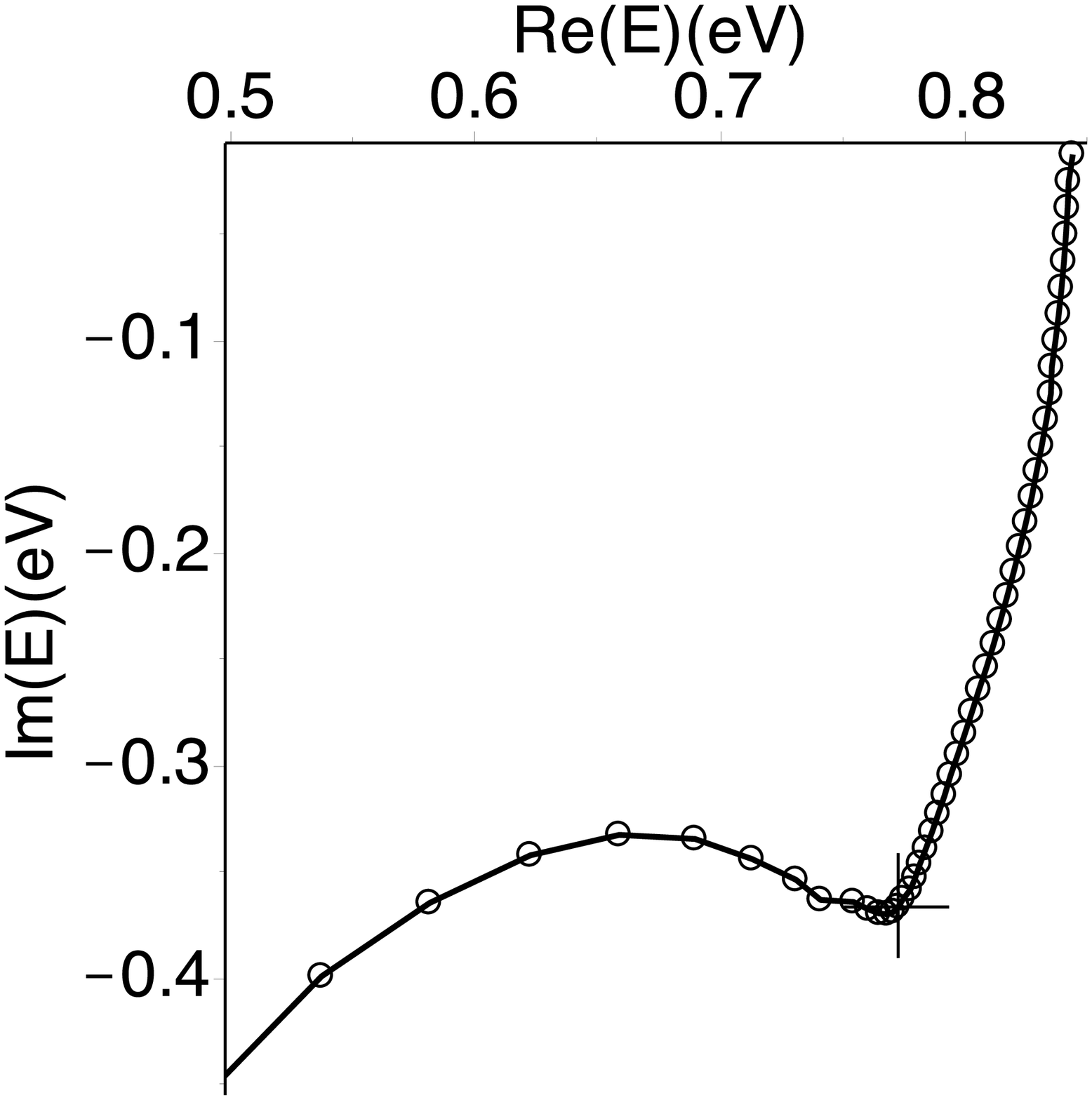}}
      \subfigure[]{\includegraphics[width=0.320\textwidth]{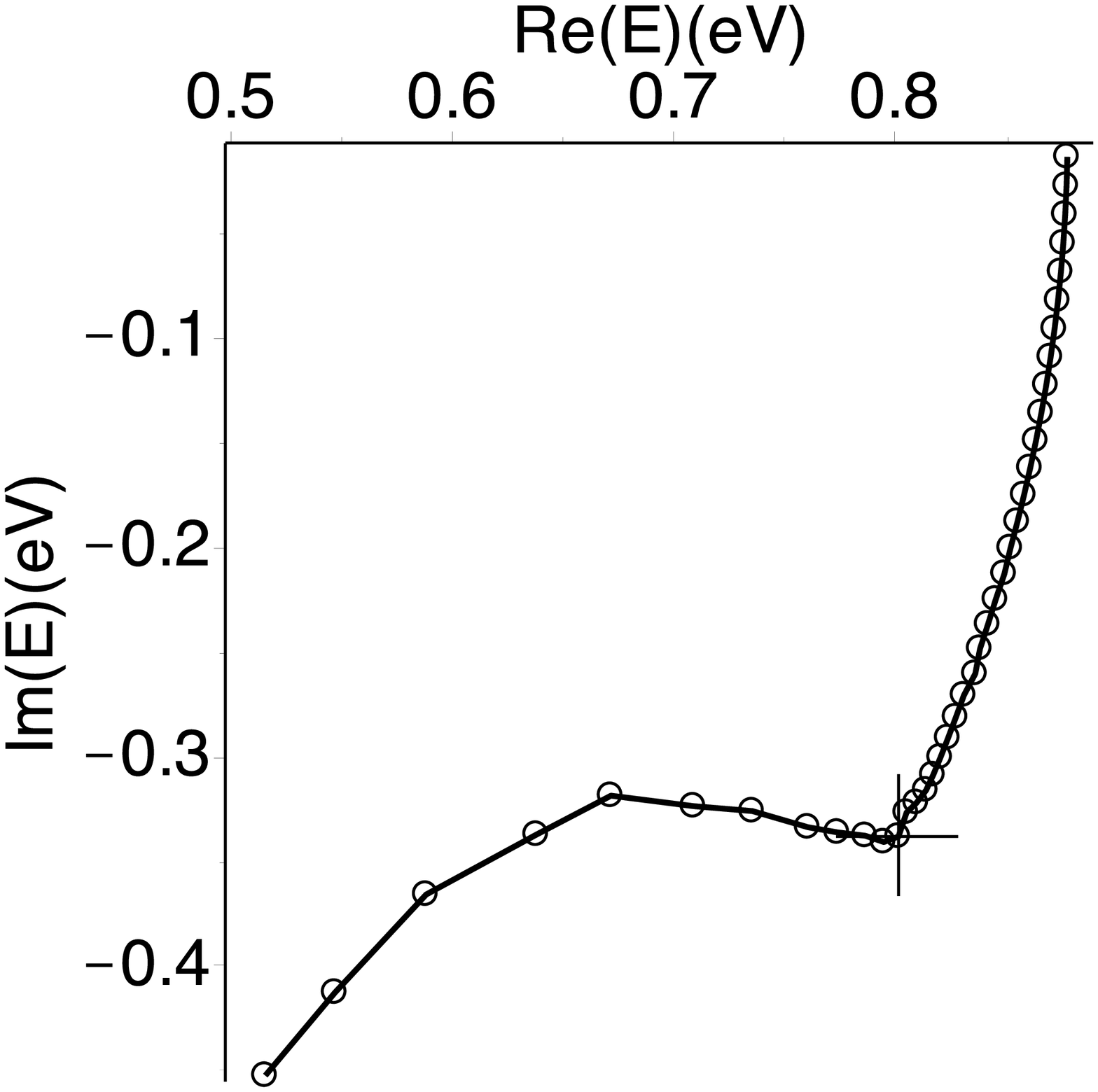}}
      \subfigure[]{\includegraphics[width=0.32\textwidth]{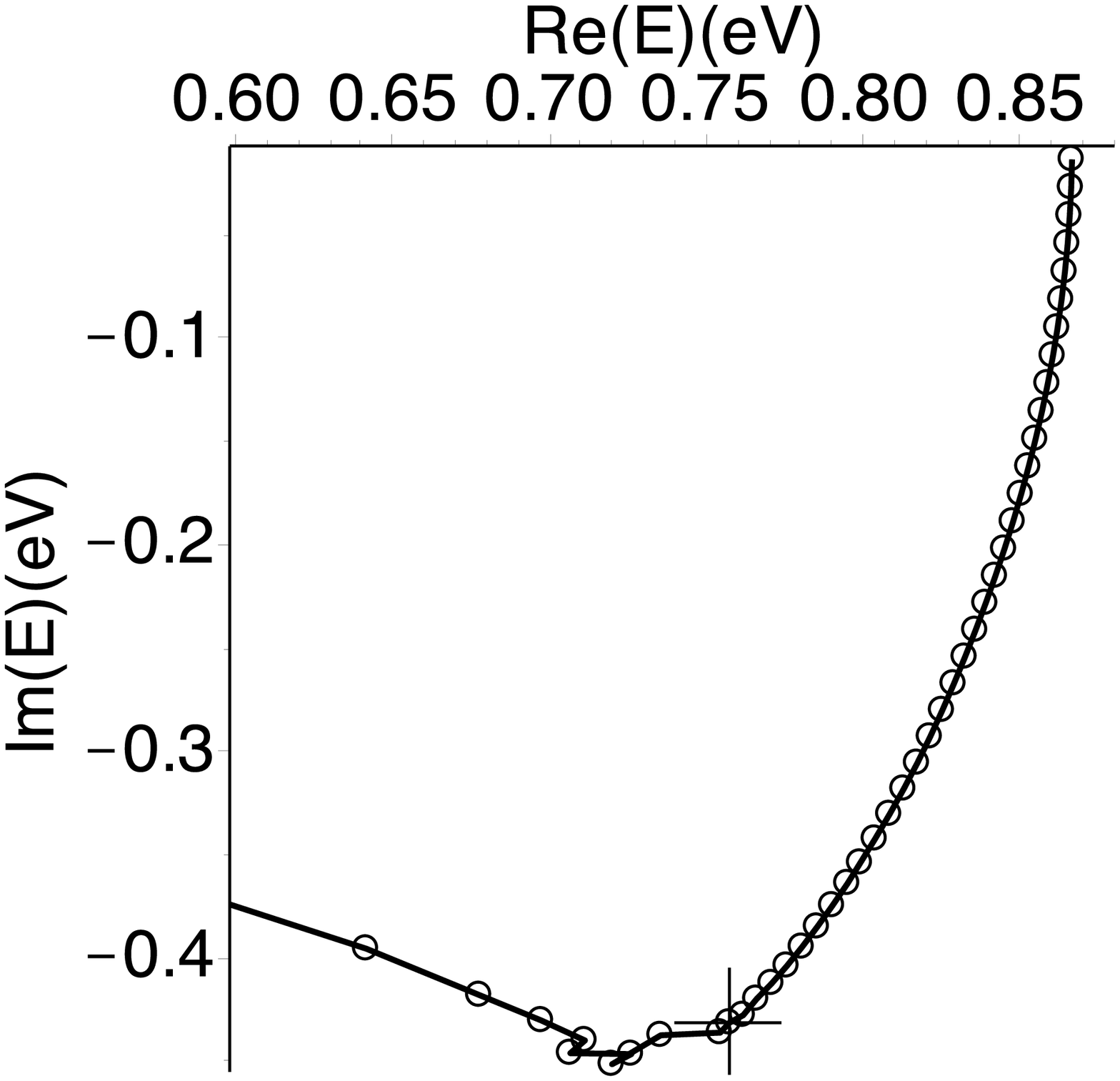}} 
      }

\caption{ The $\theta$-trajectories for  $^2P\,\mbox{Be}^{-}$ shape resonance obtained from the CMCSTEP method. The curves shown in panels (a), (b) and (c) correspond to basis sets $14s11p5d$, $14s14p2d$ and $14s14p5d$ with $2s2p3d$\,CAS, respectively, and a cross shows a  stabilized point. Computational parameters are $\alpha=$ 1\,(a), 1.03\,(b,c) and $\Delta \theta = 0.01rad$.}  
\label{fig:1}       
\end{figure*}

Although we have here presented results for resonance parameter for an atomic system, $^2P\,\mbox{Be}^{-}$, the method can be implemented for investigating shape resonance parameters for molecular systems. We had shown for the $^2\Pi_{g}\,\mbox{N}_{2}^{-}$ shape resonance \cite{Kous3} using $\mbox{M}_{1}$, and it is quite consistent with previous literature results \cite{Resc2,Meyer2,Mah3,Saj}  and experimental measurements \cite{Schultz,Som}. In the molecular case \cite{Kous3}, the CS technique for the electron-nuclear Coulomb interaction potential $-Z/|\mathbf{r}-\mathbf{R}| $ has been implemented so that $-(Z\eta^{-1}) /|\mathbf{r}-\mathbf{R}\eta^{-1}| $, where $Z$ is a nuclear charge, and $\mathbf{r}$ and $\mathbf{R}$ are  the electronic and nuclear positions relative to an origin of a fixed molecular coordinate system \cite{Donn}. We will report CMCSTEP calculations for several molecules using this procedure in the near future.

\section{Conclusions}

In this work we have developed the CMCSTEP method and presented theoretical calculations for $^2P\,\mbox{Be}^{-}$ shape resonance using three different ($\Delta$CMCSCF, $\mbox{M}_{1}$ and CMCSTEP) methods. In our group we previously developed $\Delta$CMCSCF and $\mbox{M}_{1}$ methods, however,  we here have developed and implemented for the first time the CMCSTEP method for resonance problems, using three different bases sets $14s11p5d$, $14s14p2d$ and $14s14p5d$ with $2s2p3d$\,CAS. In CMCSTEP calculations we use the CMCSCF state as an initial state. The obtained values of  $^2P\,\mbox{Be}^{-}$ shape resonance from CMCSTEP method are compared with previously obtained results in the literature. Based on our previous results from MCSTEP calculations for IPs and AEs for atomic and molecular systems, the results from CMCSTEP calculations are probably the most reliable and practical for resonance problems. 


In next step of our research work we intend to apply the CMCSTEP method to resonance problems for open-shell atomic and molecular systems, as well as studies for Feshbach and Auger resonances. 


\section*{Acknowledgments}
We thank the Robert A. Welch Foundation for financial support grant A-770. Fruitful discussions with Dr. Songbin Zhang and Dr. K. Samanta on implementing the complex version of the codes 
are gratefully acknowledged.  

\end{document}